\documentclass{article}

\usepackage{arxiv}
\usepackage{graphicx}
\usepackage{subfig}
\usepackage{makecell}
\usepackage{array,booktabs,multirow}
\usepackage{adjustbox}
\usepackage{mathtools}
\usepackage{xspace}
\usepackage{footnote}
\usepackage{afterpage}
\usepackage{tablefootnote}
\makesavenoteenv{tabular}
\makesavenoteenv{table}
\usepackage{float}
\usepackage{scrextend}
\usepackage[bottom]{footmisc}
\usepackage{amsmath}
\usepackage{stackengine}
\usepackage{scalerel}
\usepackage{yhmath}
\usepackage[T1]{fontenc}
\usepackage[linesnumbered,ruled,vlined]{algorithm2e}
\SetCommentSty{mycommfont}
\SetKwInput{KwInput}{Input}                
\SetKwInput{KwOutput}{Output}
\graphicspath{{figures/}}
\DeclareGraphicsExtensions{.pdf,.jpeg,.png}

\usepackage{url}

\begin{document}
\title{Segmentation of common and internal carotid arteries from 3D ultrasound images using adaptive triple U-Net}

\author{
Mingjie Jiang \\
  Department of Electrical Engineering\\
  City University of Hong Kong\\
  Hong Kong \\
  \And
 Yuan Zhao \\
  Department of Electrical Engineering\\
  City University of Hong Kong\\
  Hong Kong \\
   \\
  \And
  Bernard Chiu\\
  Department of Electrical Engineering\\
  City University of Hong Kong\\
  Hong Kong \\
  \texttt{bcychiu@cityu.edu.hk} \\
}
\providecommand{\keywords}[1]{\textbf{\textit{Index terms---}} #1}

\date{\today}

%

\maketitle

\begin{abstract}	
\noindent\textbf{Objective}: Vessel-wall-volume (VWV) and localized vessel-wall-thickness (VWT) measured from 3D ultrasound (US) carotid images are sensitive to anti-atherosclerotic effects of medical/dietary treatments. VWV and VWT measurements require the lumen-intima (LIB) and media-adventitia boundaries (MAB) at the common and internal carotid arteries (CCA and ICA). However, most existing segmentation techniques were capable of automating only CCA segmentation. An approach capable of segmenting the MAB and LIB from the CCA and ICA was required to accelerate VWV and VWT quantification. 

\noindent\textbf{Methods}: Segmentation for CCA and ICA were performed independently using the proposed two-channel U-Net, which was driven by a novel loss function known as the adaptive triple Dice loss (ADTL). A test-time augmentation (TTA) approach is used, in which segmentation was performed three times based on axial images and its flipped versions; the final segmentation was generated by pixel-wise majority voting.  

\noindent\textbf{Results}: Experiments involving 224 3DUS volumes produce a Dice-similarity-coefficient (DSC) of 95.1\%$\pm$4.1\% and 91.6\%$\pm$6.6\% for the MAB and LIB, in the CCA, respectively, and 94.2\%$\pm$3.3\% and 89.0\%$\pm$8.1\% for the MAB and LIB, in the ICA, respectively. TTA and ATDL independently contributed to a statistically significant improvement to all boundaries except the LIB in ICA. The total time required to segment the entire 3DUS volume (CCA+ICA) is 1.4s. 

\noindent\textbf{Conclusion}: The proposed two-channel U-Net with ADTL and TTA can segment the CCA and ICA accurately and efficiently from the 3DUS volume. 

\noindent\textbf{Significance}: Our approach has the potential to accelerate the transition of 3DUS measurements of carotid atherosclerosis to clinical research.
\end{abstract}

\section{Introduction}

Stroke is the leading cause of death in China and the second global leading cause of death \cite{who2018global,wu2019stroke}, and the prevalence of stroke has been increased by about 200\% from 1993 to 2013~\cite{wu2019stroke}. Ischemic strokes account for 87\% of all strokes \cite{chen2010ischemic}. This type of stroke is mostly caused by the blockage of a cerebral artery by an embolus. Carotid atherosclerosis is a major source of emboli, which are composed of platelet aggregates and plaque debris arising from plaque rupture. Management of atherosclerosis in high-risk population through dietary and medical therapies can reduce stroke risk by up to 80\% \cite{spence2007intensive}. Therefore, there is a critical need in developing sensitive and reproducible biomarkers for the identification of high-risk patients and monitoring plaque response to therapies.

Carotid intima-media thickness (IMT) measured from the common carotid arteries (CCA) imaged by 2D ultrasound (2DUS) is an early imaging biomarker that was shown to correlate with clinical outcomes \cite{bots1997common,o1999carotid}. However, recent investigations show that it is not a strong cardiovascular event predictor \cite{den2012common} and not sensitive to treatment effect \cite{bots2003carotid,cheng2017sensitive}. These weaknesses of IMT are attributable to several factors: (1) the small annual change of IMT ($\sim$0.015mm) does not allow the measurement of treatment effect in a clinically affordable time frame \cite{o2002intima}; (2) IMT measures vascular wall thickening, which is not directly related to atherosclerosis \cite{finn2010correlation}; (3) 2DUS requires an operator to locate an imaging plane to be scanned, making IMT measurement difficult to reproduce even for the same operator \cite{chiu2008development} These issues are addressed by the development of 3D ultrasound (3DUS) imaging.  As plaques grow/regress circumferentially as well as longitudinally and in thickness, the total plaque
volume (TPV) and vessel wall volume (VWV) measured from 3D were shown to be more
sensitive to treatment effect in several studies \cite{krasinski2009three,ainsworth20053d}. As carotid atherosclerosis is a focal disease, the evolution of the size and stability of plaques is affected by their locations in the artery \cite{zarins2004localization}. Our group has developed a methodology to measure the vessel-wall-plus-plaque thickness (VWT) on a point-by-point basis \cite{chiu2008quantification} and introduced metrics to quantify the VWT distributions \cite{cheng2017sensitive,chiu2016concise,chiu2013novel}. A VWT-based metric we introduced led to 70\% reduction in the sample size required to show treatment effect compared to VWV \cite{cheng2017sensitive}. These measurements require the segmentation of the media-adventitia (MAB) and lumen-intima (LIB) from the 3DUS image. Although these boundaries can be segmented manual as in previous clinical studies \cite{krasinski2009three,cheng2016three,cheng2017sensitive}, manual segmentation is time-consuming \cite{zhou2020voxel}, creating a bottleneck that limits the clinical utility of the 3DUS measurements. 

Table \ref{table:papers} lists previous algorithms introduced for segmenting MAB and LIB from ultrasound images. Yang et al. \cite{yang2013ultrasound} used the active shape model to perform CCA segmentation from 3DUS. Their method required manual intervention to locate nine landmark points on the contour of MAB and LIB on a slice-by-slice basis and this process is time-consuming. Ukwatta et al. \cite{ukwatta2013three} proposed a sparse level set algorithm to delineate the MAB and LIB of CCA from 3DUS. This approach requires around 2 minutes for a user to initialize the algorithm; this is an improvement compared to their earlier slice-based model, which takes around 9 minutes to initialize \cite{ukwatta2011three}. Hossain et al. \cite{hossain2015semiautomatic} proposed a stopping criterion for a level-set algorithm to segment MAB and LIB of common, internal and external carotid arteries. Their method required around 14 mins for a user to initialize the algorithm. The requirement of human interaction also leads to observer variability, which has been demonstrated in Chiu et al. \cite{chiu2013quantification} for the arteries segmented in Ukwatta et al. \cite{ukwatta2011three}. 

\begin{table*}[htbp]
\centering
\begin{adjustbox}{max width=\textwidth}
\begin{tabular}{ccccc}
\toprule[.2em]
Paper & Year & ROI  & Carotid & \# of subjects in total \\
\midrule[.2em]
Ukwatta et al. \cite{ukwatta2011three} & 2011 & Y & CCA & 21 \\
Yang et al. \cite{yang2013ultrasound} & 2013 & Y &  CCA & 17 \\
Ukwatta et al. \cite{ukwatta2013three} & 2013 & Y  & CCA & 21 \\
Menchón-Lara et al. \cite{menchon2014automatic} & 2014 & Y &  CCA & 30 \\
Hossain et al. \cite{hossain2015semiautomatic} & 2015 & Y  & CCA\&ICA\&ECA & 10 \\
Azzopardi et al. \cite{azzopardi2015carotid} & 2017 & Y  & CCA & 5 \\ 
Zhou et al. \cite{zhou2019deep} & 2019 & Y & CCA & 38 \\ 
Zhou et al. \cite{zhou2020voxel} & 2020 & Y & CCA & 1007 \\
Azzopardi et al. \cite{azzopardi2020bimodal} & 2020 & N &  CCA & 15 \\
\bottomrule[.2em]\\
\end{tabular}
\end{adjustbox}
\caption{Previous published papers on ultrasound carotid segmentation. Y and N stand for yes and no respectively. CCA, ICA and ECA stand for the common, internal and external carotid arteries. }
\label{table:papers}
\end{table*}
Deep learning algorithms have been introduced to reduce user interaction in the segmentation process. Menchón-Lara et al. \cite{menchon2014automatic} used the autoencoder and multi-layer perceptron (MLP) to detect the media-adventitia (MAB) and the lumen-intima boundaries (LIB) from 2D longitudinal ultrasound images with an ROI covering the CCA cropped manually. They applied an MLP classifier to patches centering at each pixel to be segmented. This method is limited in computational efficiency due to the use of patch-by-patch classification. Azzopardi et al. \cite{azzopardi2017automatic} developed a small U-Net-like convolutional neural network (CNN) to segment LIB and MAB using axial carotid images and the corresponding phase congruency maps. A limitation of this work is that the segmentation performance was evaluated only on five subjects. Besides, the computation of the phase congruency maps would slow down the segmentation procedure and may introduce noise into the network. The algorithm was later improved by adding the geometrically constrained terms in the loss function \cite{azzopardi2020bimodal}. These terms were shown to provide higher segmentation accuracy using leave-one-out cross-validation. However, it is questionable whether the algorithm should be evaluated by leave-one-out cross-validation. First, clinical applications could not afford a leave-one-out workflow that repeats the training procedure for each artery. Second, as each trained model was tested only on one artery, it is uncertain whether the trained model with the geometrically constrained terms can generalize to provide accurate segmentation for arteries with different shapes. Zhou et al. \cite{zhou2019deep} proposed a semi-automatic framework based on dynamic convolutional network and U-Net to segment MAB and LIB of the CCA and their method required extensive user-interaction to initialize the MAB contour. The initialization process takes 13.8 s for each artery and introduces observer variability. The same group \cite{zhou2020voxel} proposed another CCA segmentation network that integrates (1) a 3D pyramid pooling module expanded from pyramid pooling module (PPM) \cite{zhao2017pyramid} and (2) an attention mechanism to select features from the multiple upsampling paths. The results generated by the network was further smoothed by the continuous max-flow (CMF) algorithm. This method required the manual cropping of a region-of-interest (ROI), and therefore, is not fully automatic. 


Most methods introduced above requires manual identification of an ROI before the MAB and LIB segmentation. The first contribution of this work is the introduction of a CNN that does not require manual ROI identification for CCA segmentation. The proposed CCA segmentation network would save time required for ROI identification in a large clinical trial (for example, 7345 patients were followed in the stroke prevention clinic that provided images for this study \cite{spence2016determinants}). Another motivation for this work stems from the lack of automated methods to segment the internal carotid artery (ICA) from 3DUS images. VWV and VWT measurements and metrics derived from local characterization of VWT-Change from the carotid template \cite{chiu2013novel,cheng2017sensitive,krasinski2009three, chen2020development, wannarong2013progression} were shown to be sensitive in detecting effects of medical or dietary treatments in a short period of followup time and provide accurate stroke risk stratification. These 3DUS carotid measurements were made on a longitudinal coverage that incorporates the CCA and ICA. Measurements made only on CCA would be less useful as carotid plaques are much more prevalent in ICA than in CCA (the Northern Manhattan Study \cite{prabhakaran2006carotid} reported that 52\% of subjects had plaques in the ICA or the bifurcation, but only 4.3\% of subjects had plaques in the CCA). Although 3DUS measurements and metrics derived from the local characterization of VWT-Change from the carotid template \cite{chiu2013novel,cheng2017sensitive,krasinski2009three} were useful in treatment evaluation and risk stratification, these measurements have not gained widespread clinical acceptance. The major bottleneck in the workflow of VWV and VWT measurements is the requirement of manual segmentation of the MAB and LIB in the CCA and ICA. The second contribution addresses this issue by introducing a network that segments ICA with minimal user inputs. The third contribution is the introduction of the adaptive triple Dice loss function (ATDL) that involves three terms that characterize the Dice coefficients of the MAB, LIB and vessel wall (i.e., the region between the MAB and LIB). The triple Dice loss function (TDL) was first proposed for segmenting black-blood carotid MR images \cite{wu2019deep}. Our loss function differs in that the weights of the terms associated with MAB, LIB and the vessel wall were adaptively determined. Dice coefficients of the MAB and LIB are typically larger than that of the vessel wall, as the vessel wall area (and therefore, the area overlap) is much smaller than the areas covered by the MAB and LIB. The proposed loss function weighting method put a larger weight on the loss function term associated with the vessel wall adaptively and improved the segmentation performance. 

We demonstrate that the point-wise VWT on the CCA and ICA can be measured from the MAB and LIB segmented by the proposed algorithm. The distribution of VWT throughout the artery can be visualized on the carotid template previously described \cite{chen2016correspondence} and VWT-based metrics sensitive to treatment effect, such as weighted VWT average of a subject introduced in \cite{cheng2017sensitive}, can be computed directly from the carotid template. The proposed carotid segmentation framework substantially accelerates the workflow for VWV and VWT measurements, thereby promoting the clinical utility of 3D ultrasound measurements of carotid atherosclerosis.

\section{Materials and Methods}
\subsection{Image acquisition and preprocessing}

Subjects with diabetic nephropathy were recruited for the Diabetic Intervention with Vitamin to Improve Nephropathy (DIVINe) trial. They provided written informed consent to the study protocol as described previously \cite{house2010effect}. A total of 224 3DUS volumes from 56 subjects acquired at baseline and a follow-up session for both carotid arteries were available for training, validating and testing the proposed algorithm. High-resolution 3DUS images were obtained by translating an ultrasound tranducer (L12-5, Philips, Bothel, WA, USA) mounted on a mechanical assembly along the neck of the subjects for approximately 4 cm. The 2D ultrasound frames from the ultrasound machine (ATL HDI 5000, Philips, Bothel, WA, USA) were captured by a frame grabber and reconstructed into a 3D image.


Manual segmentation of the carotid arteries was used as the surrogate ground truth to train and evaluate the proposed algorithm. The CCA and ICA were manually segmented on axial planes with an interslice distance of 1 mm, as previously described \cite{egger2007validation}. The CCA was segmented for 15 mm proximal to the bifurcation and the ICA was segmented for 10 mm distal to the bifurcation. 

No manual interaction was required for identifying ROI for CCA segmentation. For ICA segmentation, a bounding box was required to be identified on two ICA axial slices in the 3DUS volume: the axial slices that are closest and furthest from the bifurcation. The rectangular ROI was defined by two manually identified landmarks at the two diagonally opposite corner of the rectangle. An ROI was automatically generated for each ICA image slice in between by linear interpolation. The bounding boxes identified for the slices closest and furthest from the bifurcation were expanded by 20 pixels before this linear interpolation to ensure that the interpolated bounding boxes enclose the ICA. As the CNN requires images of the same size to be fed in as batches, all axial slices were resliced to $256\times320$ pixels. The intensities of input images were normalized to $[0, 1]$ by linear scaling.
\subsection{Network architecture}
\begin{figure*}[htbp]
	\centering
	\includegraphics[width=\textwidth]{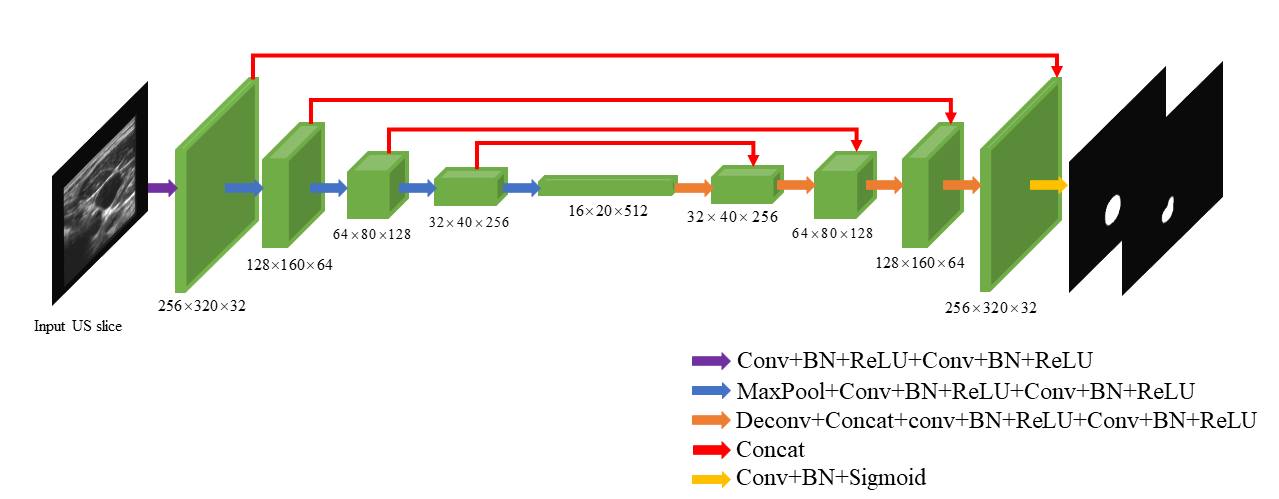}
	\caption{The architecture of the proposed two-channel CNN. The number at the bottom of each block denotes the size of the tensor. The following labels are used in the legend: Conv -- Convolution, MaxPool -- Maxpooling, BN -- Batch normalization, ReLU -- Rectified linear unit, Concat -- concatenation of two tensors, Sigmoid -- sigmoid function. \label{fig:dscn}}
\end{figure*}

Fig. \ref{fig:dscn} shows the architecture of the proposed two-channel network. The outputs in the two channels represent the probabilities that each pixel is enclosed by the MAB and LIB, respectively. The probability of each pixel being enclosed by the vessel wall was obtained by first subtracting the LIB from the MAB output channel. The result was then fed into a rectified linear unit (ReLU) $(x\mapsto\max(x, 0))$. The two-channel structure is associated with two major advantages. First, since the LIB is inside MAB, it is anticipated that features required to localize these two boundaries are similar and complementary. An integrated encoder for the two-channel input is more appropriate in this setting. Second, simultaneous segmentation of MAB and LIB by the two-channel structure is more efficient than using two independent models to perform MAB and LIB segmentations. We will compare the segmentation performance of this two-channel structure with that of the two independent U-Nets for MAB and LIB segmentations.

The U-Net in this paper employs 2D convolution of kernel size $3\times3$ with stride $1$, and uses 2D transpose convolution of kernel size $2\times2$ with stride $2$, and utilizes 2D max pooling of kernel size $2\times2$ with stride $2$. The activation function of the output layer is the sigmoid function $(x\mapsto\frac{1}{1+e^{-x}})$ and we use ReLU as activation function for other layers. In addition, batch normalization \cite{ioffe2015batch} is placed between the convolution and activation function. 
\subsection{Loss function}
\label{sc:atdl}
We propose the adaptive triple Dice loss (ATDL), which is based on the triple Dice loss (TDL) \cite{wu2019deep} but assigns weights of the three Dice losses adaptively. The triple Dice loss is designed to minimize the difference between $\widehat{y}_\text{MAB}$, $\widehat{y}_\text{LIB}$, $\widehat{y}_\text{wall}$ and their corresponding ground truth simultaneously, and TDL is defined as follows:
\begin{equation*}
    \begin{split}
        \mathcal{L} &= \alpha\mathcal{L}_\text{DSC}(\widehat{y}_\text{MAB}, y_\text{MAB})+\beta\mathcal{L}_\text{DSC}(\widehat{y}_\text{LIB}, y_\text{LIB})+ \gamma\mathcal{L}_\text{DSC}(\widehat{y}_\text{CVW}, y_\text{CVW}),
    \end{split}
\end{equation*}
where $y_\text{MAB}$, $y_\text{LIB}$, $y_\text{CVW}$ are binary images denoting whether each pixel is enclosed by the manually segmented MAB, LIB and the vessel wall boundary (CVW), respectively, with $y_\text{CVW}=y_\text{MAB}\backslash y_\text{LIB}$. $\widehat{y}_\text{MAB}$, $\widehat{y}_\text{LIB}$, $\widehat{y}_\text{CVW}$ represent the probabililty of each pixel being enclosed by the MAB, LIB and CVW, respectively, as determined by the algorithm. $\alpha$, $\beta$ and $\gamma$ are the hyperparameters used to balance the importance of three Dice losses. $\mathcal{L}_\text{DSC}(y,\widehat{y}) = 1 - \frac{2\left|y\times\widehat{y}\right|}{\left|y+\widehat{y}\right|}$, where $+$ and $\times$ represent pixel-by-pixel addition and multiplication, respectively. $\alpha+\beta+\gamma=1$. TDL assigns the same weight to the three Dice losses (i.e., $\alpha=\beta=\gamma=\frac{1}{3}$).

$\mathcal{L}_\text{DSC}(\widehat{y}_\text{CVW}, y_\text{CVW})$ is larger than $\mathcal{L}_\text{DSC}(\widehat{y}_\text{MAB}, y_\text{MAB})$ and $\mathcal{L}_\text{DSC}(\widehat{y}_\text{LIB}, y_\text{LIB})$ because the vessel wall area is much smaller than the areas covered by MAB and LIB. In the course of the training phase, $\mathcal{L}_\text{DSC}(\widehat{y}_\text{MAB}, y_\text{MAB})$ and $\mathcal{L}_\text{DSC}(\widehat{y}_\text{LIB}, y_\text{LIB})$ reduce quickly to a value that is substantially lower than $\mathcal{L}_\text{DSC}(\widehat{y}_\text{CVW}, y_\text{CVW})$. Instead of using a uniform weight for the three loss terms in the entire training process, we propose to focus more on reducing $\mathcal{L}_\text{DSC}(\widehat{y}_\text{CVW}, y_\text{CVW})$ at the later stage of the training process. A workflow is proposed in which $\alpha$ and $\beta$ are controlled adaptively by $\mathcal{L}_\text{DSC}(\widehat{y}_\text{MAB}, y_\text{MAB})$ and $\mathcal{L}_\text{DSC}(\widehat{y}_\text{LIB}, y_\text{LIB})$, respectively. When these two loss functions are already small in the middle of the training phase, $\alpha$ and $\beta$ are reduced, allowing more emphasis to be placed on minimizing $\mathcal{L}_\text{DSC}(\widehat{y}_\text{CVW}, y_\text{CVW})$.

The weights for the three losses evolve adaptively in ATDL according to the following equations: 
\begin{equation}
\label{eq:atdl}
    \begin{split}
        \alpha^*(\mathcal{L}_1)&=\frac{\mathcal{L}_1}{3\bigg(1+a(1-\mathcal{L}_1)\bigg)}\\
        \beta^*(\mathcal{L}_2)&=\frac{\mathcal{L}_2}{3\bigg(1+a(1-\mathcal{L}_2)\bigg)}\\
        \gamma^*(\mathcal{L}_1, \mathcal{L}_2)&=1-\alpha^*(\mathcal{L}_1)-\beta^*(\mathcal{L}_2),
    \end{split}
\end{equation}
where $a>0$ is a hyperparamter. Note that $\alpha^*(0)=\beta^*(0)=0$, $\alpha^*(1)=\beta^*(1)=\frac{1}{3}$. In addition, $\alpha^*$ and $\beta^*$ are monotonic functions increasing from $0$ to $\frac{1}{3}$ when $\mathcal{L}_1$ and $\mathcal{L}_2$ increase from $0$ to $1$.

The new algorithm, which we call the adaptive triple dice loss (ATDL) approach, consists of two phases. A uniform weight was used in the first half and an adaptive weight defined in Eq. \ref{eq:atdl} was used in the second half. As $\mathcal{L}_\text{DSC}(\widehat{y}_\text{MAB}, y_\text{MAB})$ and $\mathcal{L}_\text{DSC}(\widehat{y}_\text{LIB}, y_\text{LIB})$ are relatively large at the beginning of the training, a uniform emphasis was given to MAB, LIB and CVW segmentation to facilitate the localization of MAB and LIB. When $\mathcal{L}_\text{DSC}(\widehat{y}_\text{MAB}, y_\text{MAB})$ and $\mathcal{L}_\text{DSC}(\widehat{y}_\text{LIB}, y_\text{LIB})$ are small, more emphasis was given to CVW segmentation according to Eq. \ref{eq:atdl} to refine the carotid segmentation.

\subsection{Data augmentation in training and testing}
The training set was augmented by reconstructing MAB and LIB surfaces from manual segmentation using a shape-based interpolation method \cite{grevera1996shape} and then reslicing the image and reconstructed surface with an inter-slice distance of 0.1 mm. The training set generated this way was only involved in training the proposed network and not involved in the testing phase. Only axial slices with manual segmentation were involved in the evaluation of the network. A total of 53,968 axial image slices with manual or interpolated contours were generated by this procedure, in which 33,824 are CCA slices and 20,044 are ICA slices. 

The following geometric operations were performed to further augment the training set: (a) Flipping: An image is randomly flipped horizontally about the image center with a probability $P_1$ and then is randomly flipped vertically with a probability $P_2$. In this study, $P_1=P_2=0.5$. (b) Translation: An image was randomly translated along x- and y-axes by $T_x$ and $T_y$ pixels, where $T_x$ and $T_y$ are uniformly sampled from $[0, 0.2H]$ and $[0, 0.2W]$, where $H$ and $W$ are the height and the width of the image. (c) Rotation: An image was randomly rotated by $d$ degrees about the image center, where $d$ is uniformly sampled from $[-20,20]$.

\begin{figure*}[htbp]
	\centering
	\includegraphics[width=\textwidth]{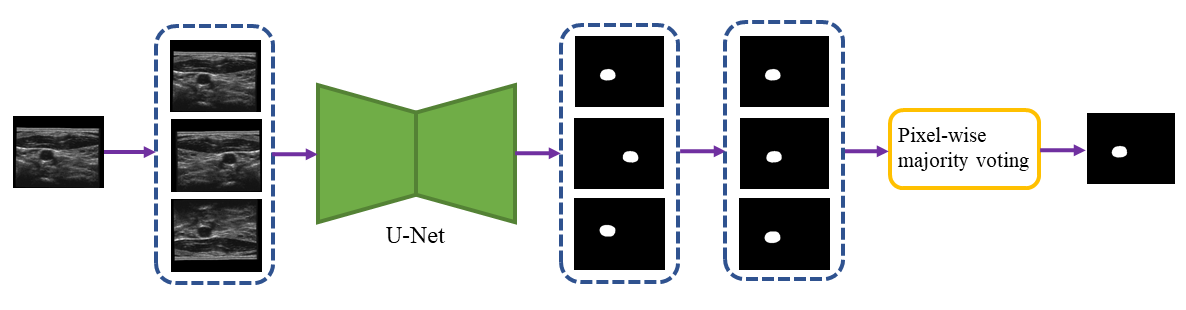}
	\caption{Workflow of test time augmentation (TTA) procedure\label{fig:TTA}}
\end{figure*}

We performed image flipping in the inference stage to improve the segmentation accuracy. Fig. \ref{fig:TTA} shows the workflow of this scheme. In particular, each 2D axial carotid image was flipped horizontally and vertically, resulting in three versions of the same image (i.e., the original, the horizontally and vertically flipped images). These three versions of the same image were fed into the network to generate three different segmentation results. The segmentation results generated for the flipped images were flipped back to the original position, resulting in three aligned segmentation results for the same image slice. Pixel-wise majority voting was performed to generate the final segmentation result for the 2D axial image.

\section{Experiments}

\subsection{Experimental settings and implementations}
Three major components were integrated in the proposed network: (i) the novel ATDL function minimized in the two-channel U-Net architecture; (ii) test-time augmentation (TTA) and pixel-wise majority voting; (iii) training set augmentation by MAB and LIB surface reconstruction and reslicing. Experiments were designed to evaluate each of the three components independently. In addition, as the effects of the two-channel architecture and the triple Dice loss have not been evaluated for carotid ultrasound images, we also evaluated the effects of these two components in this study. Specifically, four models were evaluated, which we call single Dice loss (SDL), double Dice loss (DDL), triple Dice loss (TDL) and ATDL. The SDL setting involves training two independent single-channel U-Nets. In this setting, the U-Nets used for MAB and LIB segmentations were trained separately with $\mathcal{L}_\text{DSC}(\widehat{y}_\text{MAB}, y_\text{MAB})$ and $\mathcal{L}_\text{DSC}(\widehat{y}_\text{LIB}, y_\text{LIB})$, respectively. The DDL and TDL settings involve training the two-channel U-Nets with the loss functions $1/2\mathcal{L}_\text{DSC}(\widehat{y}_\text{MAB}, y_\text{MAB})+ 1/2\mathcal{L}_\text{DSC}(\widehat{y}_\text{LIB}, y_\text{LIB})$, $1/3\mathcal{L}_\text{DSC}(\widehat{y}_\text{MAB}, y_\text{MAB})+1/3\mathcal{L}_\text{DSC}(\widehat{y}_\text{LIB}, y_\text{LIB})+1/3\mathcal{L}_\text{DSC}(\widehat{y}_\text{CVW}, y_\text{CVW})$, respectively. The training workflow for the ATDL setting was described in Section \ref{sc:atdl}. In each of the four configurations (i.e., SDL, DDL, TDL and ATDL), the improvement attributable to TTA was independently assessed by comparing the segmentation accuracy of each model with and without TTA (denoted by <Model> vs. <Model> + TTA, where <Model> = SDL, DDL, TDL or ATDL). The improvement attributable to training set augmentation by MAB and LIB surface reconstruction/reslicing was evaluated by comparing the segmentation performance produced by the ATDL + TTA model with and without this training set augmentation.

We randomly selected 60\% (n = 136), 20\% (n = 44) and 20\% (n = 44) 3DUS images for CNN training, validation and testing, respectively. Images acquired for the two carotid arteries of a single patient are included in the same partition. 

Python 3.7 was used as the programming language. Keras \cite{chollet2015keras} with TensorFlow \cite{tensorflow2015-whitepaper} backend was used as the deep learning framework. Experiments were performed on a computer with an Intel Xeon Silver 4108 CPU, 30GB memory and a Graphics Processing Unit (GPU) of Nvidia GeForce RTX 2080 Ti with 11GB frame buffer. 

The stochastic optimization algorithm Adam was employed to train the networks. The training hyperparameters are given as follows: adaptive parameter $a=0.5$, learning rate $\alpha=10^{-3}$, momentum parameters $\beta_1=0.9$, $\beta_2=0.999$, batch size is $8$, number of epochs is $50$.

\subsection{Evaluation metrics and statistical analysis}
The MAB and LIB probability maps output by the proposed two-channel network were binarized using a threshold of 0.5. Since the probability was either very close to 0 or 1, the segmentation results did not vary for a large range of thresholding probabilities. The boundaries generated in this way were evaluated by region-, distance- and volume-based metrics describe below. The Dice similarity coefficient (DSC) was used to evaluate the area overlap between the algorithm and manual segmentation: 
\begin{equation*}
\begin{split}
\text{DSC}&= 2\frac{\left|A\cap M\right|}{\left|A\right|+\left|M\right|}
\end{split}
\end{equation*}
where $M$ and $A$ denote the regions enclosed by the manual and algorithm segmentation, respectively, and $\left|\cdot\right|$ denotes the area of the operand.

Distance-based metrics evaluate the distance between the manual and algorithm segmentation. The manually segmented and the algorithm generated contours were matched on a point-by-point basis using the symmetric correspondence algorithm \cite{chiu2008quantification}. Distances between each pair of corresponding points are denoted by $\{d_i\}_{i=1}^K$, where $K$ is the total number of corresponding pairs. Two distance-based metrics, known as the mean absolute difference (MAD) and maximum absolute difference (MAXD) were defined below to summarize the distance measurements:   
\begin{equation*}
\begin{split}
\text{MAD}&=\frac{1}{K}\sum_{i=1}^Kd_i\\
\text{MAXD}&=\max_{i=1,...,K}\{d_i\},
\end{split}
\end{equation*}

Manual measurement of VWV was shown to be sensitive to treatment effect \cite{krasinski2009three}. A major motivation for the development of the proposed algorithm is to accelerate the VWV measurement workflow. Therefore, a comparison between the VWV measured from manual and algorithm-generated segmentations is important in assessing the clinical utility of the algorithm. Two-dimensional contours segmented from axial image slices were reconstructed to three-dimensional MAB and LIB surfaces using a previously proposed surface reconstruction algorithm \cite{chiu2008quantification}. The correlation and agreement between VWV measurements obtained from manual segmentation and the proposed ATDL approach were assessed using the Pearson's correlation coefficient (r) and the Bland-Altman analysis, respectively. Statistical analyses were performed using GraphPad Prism version 7.1.3 for Windows (GraphPad Software, San Diego, CA) and Matlab R2020a (The Mathworks, Inc., Natick, MA). Tukey's tests \cite{tukey1949comparing} were performed to compare metrics produced by different experimental settings. Tukey’s test is similar to the t-test, except that it corrects for the
family-wise type I error rate \cite{walpole1993probability}.

\section{Experimental results}
\label{sec:exp}


\begin{table*}[tbp]
\centering
\begin{adjustbox}{max width=\textwidth}
\begin{tabular}{cccccccc}
\toprule[.1em]
\multirow{2}{*}{carotid} & \multirow{2}{*}{model} & \multicolumn{2}{c}{$\overline{DSC}$} & \multicolumn{2}{c}{$\overline{MAD}$} & \multicolumn{2}{c}{$\overline{MAXD}$}\\
& & MAB & LIB & MAB & LIB & MAB & LIB \\
\midrule[.1em]
    \multirow{8}{*}{CCA} & SDL & {$0.919\pm0.107$}    &{$0.876\pm0.100$}  & {$2.047\pm3.129$}& {$2.213\pm1.866$}    &{$6.893\pm11.874$} & {$7.122\pm7.030$}\\
    & SDL$+$TTA & {$0.924\pm0.099$}    &{$0.882\pm0.112$}  & {$1.921\pm2.814$}& {$2.145\pm2.460$}    &{$6.551\pm10.802$} & {$6.844\pm7.197$}\\
    & DDL & {$0.927\pm0.112$}    &{$0.887\pm0.128$}  & {$1.789\pm2.230$}& {$1.998\pm1.503$}    &{$6.015\pm7.189$} & {$5.934\pm5.282$}\\ 
& DDL$+$TTA & {$0.932\pm0.119$}    &{$0.889\pm0.129$}  & {$1.663\pm2.449$}& {$1.984\pm1.648$}    &{$5.516\pm6.847$} & {$5.882\pm5.499$}\\    
  & TDL & {$0.928\pm0.088$}    &{$0.887\pm0.102$}  & {$1.822\pm2.582$}& {$1.916\pm1.327$}    &{$5.858\pm7.010$} & {$6.255\pm5.513$}\\
  
    & TDL$+$TTA & {$0.940\pm0.065$}    &{$0.899\pm0.098$}  & {$1.568\pm1.644$}& {$1.907\pm2.291$}    &{$5.261\pm5.990$} & {$6.032\pm5.980$}\\
& ATDL&  {$0.948\pm0.055$}    & {$0.913\pm0.072$}  &  {$1.418\pm1.822$}&  {$1.626\pm1.166$}    & {$4.771\pm5.950$} &{$5.501\pm5.156$}\\
& ATDL$+$TTA & \boldmath {$0.951\pm0.041$}    &\boldmath {$0.916\pm0.066$}  & \boldmath {$1.362\pm1.546$}& \boldmath {$1.595\pm1.217$}    &\boldmath {$4.641\pm5.692$} &\boldmath {$5.394\pm5.238$}\\
& {ATDL$+$TTA w/o surface reslicing}  & {$0.943\pm0.068$}    &{$0.904\pm0.100$}  & {$1.783\pm3.703$}& {$1.885\pm3.007$}    &{$5.461\pm8.421$} & {$5.898\pm7.149$}\\
\midrule[.1em]
\multirow{8}{*}{ICA} & SDL  & {$0.916\pm0.039$}    &{$0.860\pm0.143$}  & {$1.775\pm1.000$}& {$2.171\pm5.170$}    &{$4.757\pm2.658$} & {$5.501\pm7.796$}\\
& SDL$+$TTA  & {$0.928\pm0.033$}    &{$0.873\pm0.119$}  & {$1.503\pm0.781$}& {$2.197\pm6.205$}    &{$4.254\pm2.331$} & {$5.334\pm9.108$}\\
& DDL  & {$0.927\pm0.060$}    &{$0.873\pm0.137$}  & {$1.530\pm1.272$}& {$1.710\pm1.645$}    &{$4.590\pm3.119$} & {$4.622\pm4.021$}\\
& DDL$+$TTA  & {$0.930\pm0.052$}    &{$0.881\pm0.118$}  & {$1.449\pm1.092$}& {$1.615\pm1.405$}    &{$4.390\pm2.902$} & {$4.462\pm3.677$}\\
& TDL & {$0.928\pm0.063$}    &{$0.884\pm0.102$}  & {$1.571\pm1.001$}& {$1.607\pm1.217$}    &{$4.247\pm2.787$} & {$4.418\pm3.497$}\\
& TDL$+$TTA  & {$0.931\pm0.043$}    &{$0.889\pm0.097$}  & {$1.463\pm0.894$}& {$1.544\pm1.232$}    &{$3.979\pm2.390$} & {$4.255\pm3.416$}\\
 & ATDL  &  {$0.938\pm0.035$}    &  {$0.884\pm0.081$}  &  {$1.346\pm1.576$}&  {$1.615\pm1.251$}    &  {$3.891\pm3.184$} &  {$4.351\pm3.399$}\\
& ATDL$+$TTA  & \boldmath {$0.942\pm0.033$}    & \boldmath {$0.890\pm0.081$}  & \boldmath {$1.290\pm1.557$}& \boldmath {$1.544\pm1.152$}    & \boldmath {$3.689\pm3.015$} & \boldmath {$4.193\pm3.220$}\\
& ATDL$+$TTA w/o surface reslicing  & {$0.911\pm0.047$}    &{$0.886\pm0.096$}  & {$1.978\pm1.159$}& {$1.677\pm1.409$}    &{$5.129\pm2.452$} & {$4.524\pm3.686$}\\
\bottomrule[.1em]\\
\end{tabular}
\end{adjustbox}
\caption{The means and the standard deviations of Dice coefficient (DSC), mean absolute difference (MAD) and maximum absolute difference (MAXD) in MAB and LIB segmentations attained in all experimental settings. }
\label{table:cca_ica}
\end{table*}

Table \ref{table:cca_ica} shows the distance- and area-based metrics for nine experimental settings (4 loss function settings with/without test time augmentation (TTA), ADTL + TTA trained without data augmentation by MAB and LIB surface reconstruction/reslicing). All metrics indicate that the ADTL + TTA setting produces the highest accuracy among all settings for CCA and ICA segmentations. Table \ref{table:tukey} shows the p-values of Tukey's tests performed for 10 pairs of settings. The first four pairings in Table \ref{table:tukey} evaluates the contribution of the TTA strategy. The fifth and the sixth pairings evaluate the effects of the two-channel U-Net and TDL, respectively. The next three pairings evaluate the contribution of ATDL with respect to SDL, DDL and TDL. The last pairing evaluates the effect of our surface reslicing data augmentation strategy. Results in Table \ref{table:tukey} show that TTA has a significant effect in most pairings for CCA and ICA segmentations. The data augmentation strategy involving surface reconstruction and reslicing has a significant effect, except in the LIB of ICA. The two-channel U-Net and the introduction of the TDL function, proposed in Wu et al. \cite{wu2019deep} for segmentation in MRI carotid images, do not contribute independently to statistically significant improvement, as shown in the p-values obtained in the (SDL+TTA vs. DDL+TTA) and the (DDL+TTA vs. TDL+TTA) comparisons for CCA and ICA segmentations. 

\begin{figure*}[tbp]
	\centering
	\renewcommand{\thesubfigure}{1a}
\subfloat[(0.754, 0.777)]{
\includegraphics[width=0.16\textwidth]{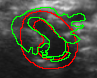}}
\renewcommand{\thesubfigure}{2a}
\subfloat[(0.832, 0.807)]{
\includegraphics[width=0.16\textwidth]{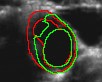}}
\renewcommand{\thesubfigure}{3a}
\subfloat[(0.408, 0.469)]{
\includegraphics[width=0.16\textwidth]{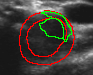}}
\renewcommand{\thesubfigure}{4a}
\subfloat[(0.680, 0.603)]{
\includegraphics[width=0.16\textwidth]{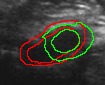}}
\renewcommand{\thesubfigure}{5a}
\subfloat[(0.803, 0.839)]{
\includegraphics[width=0.16\textwidth]{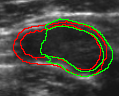}}\\
	\renewcommand{\thesubfigure}{1b}
\subfloat[(0.913, 0.906)]{
\includegraphics[width=0.16\textwidth]{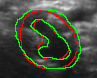}}
\renewcommand{\thesubfigure}{2b}
\subfloat[(0.959, 0.958)]{
\includegraphics[width=0.16\textwidth]{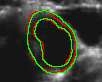}}
\renewcommand{\thesubfigure}{3b}
\subfloat[(0.925, 0.934)]{
\includegraphics[width=0.16\textwidth]{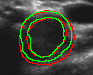}}
\renewcommand{\thesubfigure}{4b}
\subfloat[(0.840, 0.867)]{
\includegraphics[width=0.16\textwidth]{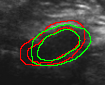}}
\renewcommand{\thesubfigure}{5b}
\subfloat[(0.905, 0.934)]{
\includegraphics[width=0.16\textwidth]{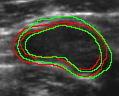}}\\
	\caption{The effect of ATDL demonstrated in five example images. The green boundaries represent algorithm-generated boundaries and the red boundaries represent manually segmented boundaries. (1a)-(5a) show the boundaries segmented in the TDL setting and (1b)-(5b) show the boundaries segmented using the ATDL approach. The tuple below each image represents the DSC for MAB and LIB segmentations. \label{fig:examples}}
\end{figure*}

\begin{figure*}[htbp]
\centering
\renewcommand{\thesubfigure}{}
\subfloat[]{
\includegraphics[width=0.1\textwidth]{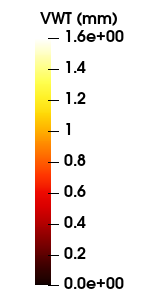}}
\renewcommand{\thesubfigure}{(1a)}
\subfloat[]{
\includegraphics[width=0.4\textwidth]{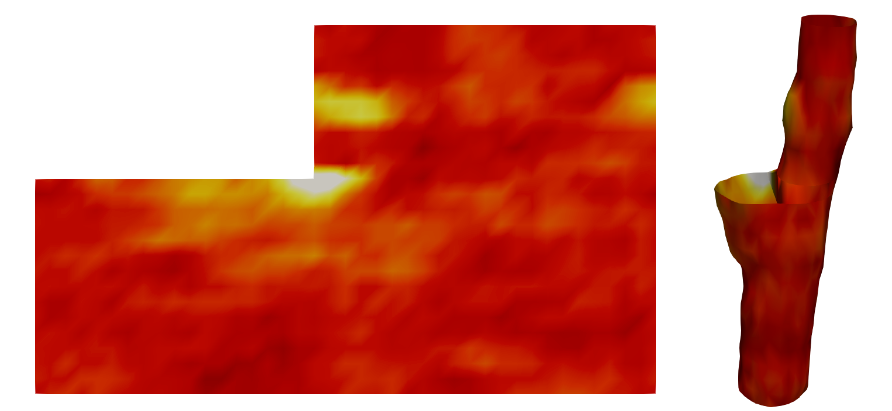}}
\renewcommand{\thesubfigure}{(1b)}
\subfloat[]{
\includegraphics[width=0.4\textwidth]{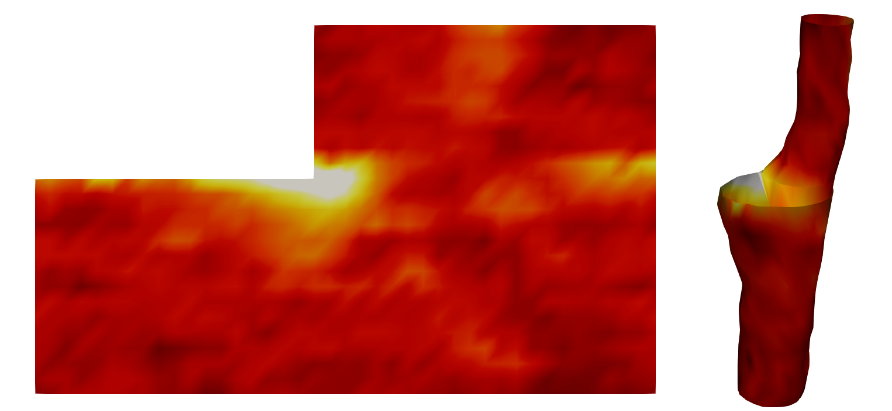}}
\\
\renewcommand{\thesubfigure}{}
\subfloat[]{
\includegraphics[width=0.1\textwidth]{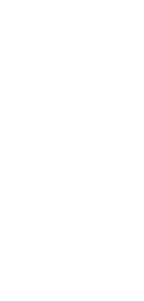}}
\renewcommand{\thesubfigure}{(2a)}
\subfloat[]{
\includegraphics[width=0.4\textwidth]{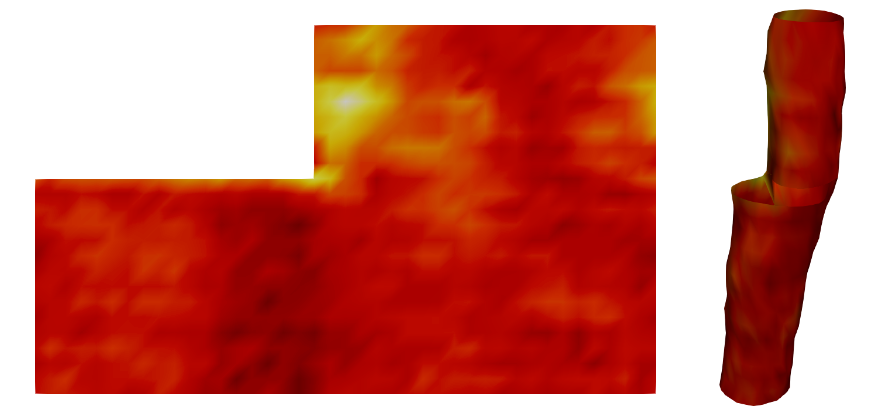}}
\renewcommand{\thesubfigure}{(2b)}
\subfloat[]{
\includegraphics[width=0.4\textwidth]{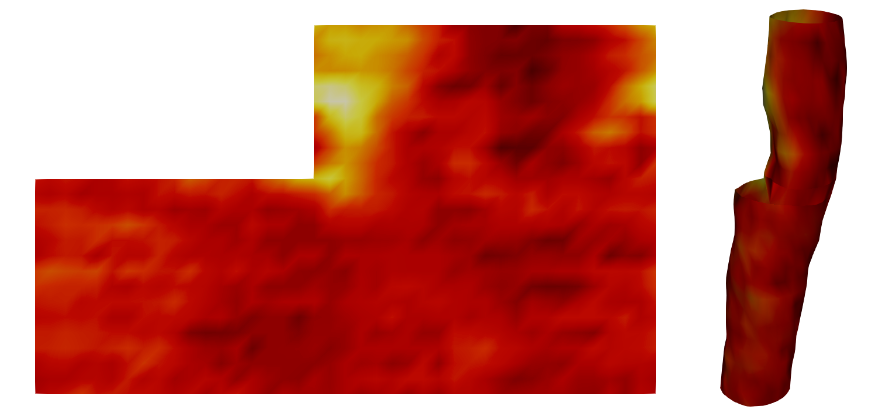}}
\\
\renewcommand{\thesubfigure}{}
\subfloat[]{
\includegraphics[width=0.1\textwidth]{placeholder.png}}
\renewcommand{\thesubfigure}{(3a)}
\subfloat[]{
\includegraphics[width=0.41\textwidth]{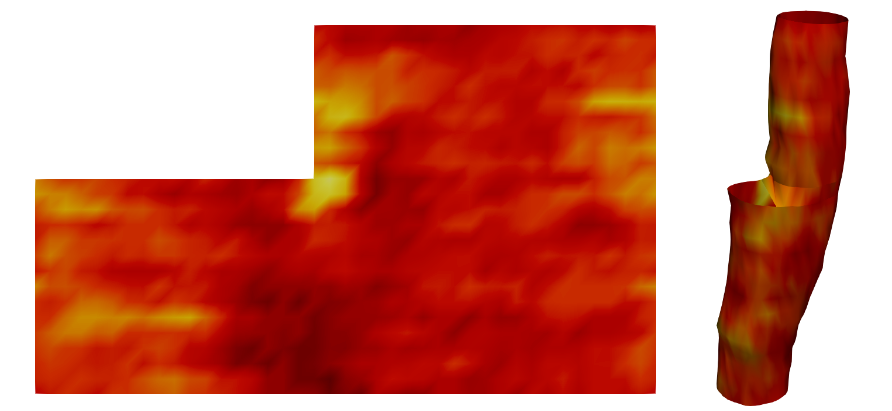}}
\renewcommand{\thesubfigure}{(3b)}
\subfloat[]{
\includegraphics[width=0.4\textwidth]{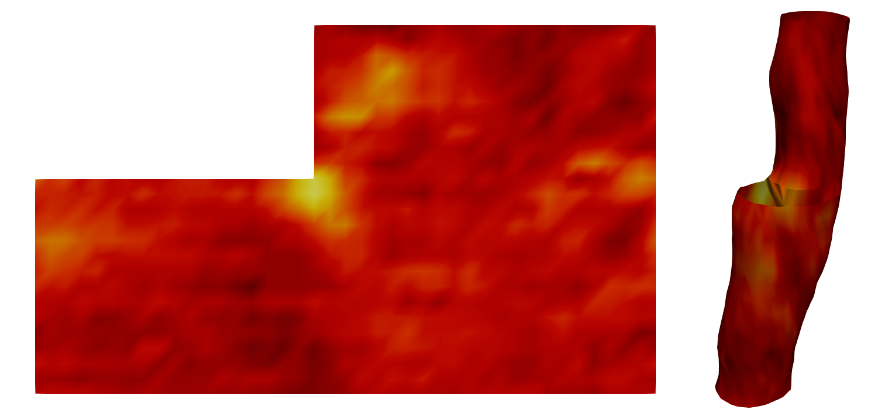}}
\caption{Flattened vessel wall thickness map and its corresponding 3D vessel wall thickness surface for three example arteries. Each row shows the maps generated by the same artery. (1a), (2a) and (3a) show the maps generated using algorithm-generated boundaries, and (1b), (2b) and (3b) show the maps generated using manually segmented boundaries.  \label{fig:VWT}}
\end{figure*}

In contrast, the proposed ATDL approach produced a significantly higher DSC for CCA and ICA segmentations than the TDL approach as shown in Table \ref{table:tukey}. The improvement attributable to the proposed ATDL approach is shown in Fig. \ref{fig:examples}, which shows the MAB and LIB contours for example cases from five arteries in the testing set. In these examples, the MAB and LIB contours segmented by the TDL model missed large sections of the carotid artery and the segmentation inaccuracy was largely corrected by the proposed ATDL model. Fig. \ref{fig:VWT} compares the 2D and 3D VWT maps generated manually and using the ADTL + TTL model with training data augmentation by surface reslicing. The 2D flattened maps were generated by cutting and unfolding the VWT maps, as previously described \cite{chen2016correspondence}. The result suggests that VWT maps can be generated accurately based on the proposed algorithm. The availability of the proposed segmentation tool will substantially accelerate the workflow of clinical evaluation of the effect of anti-atherosclerotic treatments, such as those described in our previous clinical studies \cite{krasinski2009three, cheng2017sensitive}.

\begin{table*}[tbp]
\centering
\begin{adjustbox}{max width=\textwidth}
\begin{tabular}{cccccc}
\toprule[.1em]
\multirow{2}{*}{Carotid} & \multirow{2}{*}{Effect} &
\multirow{2}{*}{Setting 1} & \multirow{2}{*}{Setting 2} & \multicolumn{2}{c}{$DSC$} \\
& & & & MAB & LIB \\
\midrule[.1em]
\multirow{11}{*}{CCA} & \multirow{4}{*}{Effect of TTA} & SDL & SDL$+$TTA & $0.0019$ & $0.0022$ \\
& & DDL & DDL$+$TTA & $0.1205$ & $0.4113$ \\
&  & TDL & TDL$+$TTA & $<0.0001$& $<0.0001$ \\
&   & ATDL & ATDL$+$TTA & $0.0044$& $0.0042$ \\
\cmidrule{2-6}
& \makecell[c]{Advantage of two-\\channel U-Net} & SDL$+$TTA & DDL$+$TTA & $0.6517$ & $0.7224$ \\
\cmidrule{2-6}
& Effect of TDL & DDL$+$TTA & TDL$+$TTA & $0.4594$ & $0.4272$ \\
\cmidrule{2-6}
& \multirow{3}{*}{\makecell[c]{Advantage of ATDL}} & ATDL$+$TTA & SDL$+$TTA & $<0.0001$ & $<0.0001$ \\
& & ATDL$+$TTA & DDL$+$TTA & $0.0001$ & $<0.0001$ \\
& & ATDL$+$TTA & TDL$+$TTA & $<0.0001$ &$<0.0001$\\
\cmidrule{2-6}
&  \makecell[c]{Effect of data augmentation\\ by surface reslicing}&\makecell[c]{ATDL$+$TTA\\w/o surface reslicing} & ATDL$+$TTA & $0.0011$ &$0.0003$ \\
\midrule[.1em]
\multirow{10}{*}{ICA} & \multirow{4}{*}{Effect of TTA} & SDL & SDL$+$TTA & $<0.0001$&$0.0122$ \\
& & DDL & DDL$+$TTA & $<0.0001$&$<0.0001$ \\
&  & TDL & TDL$+$TTA & $0.0143$&$<0.0001$ \\
&   & ATDL & ATDL$+$TTA & $<0.0001$&$0.2567$ \\
\cmidrule{2-6}
& \makecell[c]{Advantage of two-\\channel U-Net} & SDL$+$TTA & DDL$+$TTA & $0.9524$ & $0.5876$ \\
\cmidrule{2-6}
& Effect of TDL & DDL$+$TTA & TDL$+$TTA & $>0.9999$ & $0.2665$\\
\cmidrule{2-6}
& \multirow{3}{*}{\makecell[c]{Advantage of ATDL}} & ATDL$+$TTA & SDL$+$TTA & $<0.0001$ &$0.0007$ \\
& & ATDL$+$TTA & DDL$+$TTA & $<0.0001$ &$0.3790$ \\
& & ATDL$+$TTA & TDL$+$TTA & $<0.0001$ &$>0.9999$ \\
\cmidrule{2-6}
&  \makecell[c]{Effect of data augmentation\\ by surface reslicing} & \makecell[c]{ATDL$+$TTA\\w/o surface reslicing} & ATDL$+$TTA & $<0.0001$ &$0.7749$  \\
\bottomrule[.1em]\\
\end{tabular}
\end{adjustbox}
\caption{P-values of Tukey's test performed to evaluate individual contributions attributable to (1) the test-time augmentation unit, (2) two-channel U-Net, (3) the use of triple Dice loss (TDL), (4) the use of the adaptive triple Dice loss (ATDL) and (5) data augmentation by reslicing. }
\label{table:tukey}
\end{table*}
\begin{table}[t]
\centering
\begin{adjustbox}{max width=\textwidth}
\begin{tabular}{cccc}
\toprule[.1em]
carotid & model & time (s/epoch) & total time (h)\\
\midrule[.1em]
\multirow{4}{*}{CCA} &SDL & 931.47 & 12.94\\
&DDL & 491.34 & 6.82\\ 
&TDL & 498.14& 6.92 \\ 
&ATDL & 501.02& 6.96\\
\midrule[.1em]
\multirow{4}{*}{ICA} &SDL & 606.02& 8.42\\
&DDL & 309.93&4.30\\
&TDL & 304.96&4.24\\
&ATDL & 305.38&4.24\\
\bottomrule[.1em]\\
\end{tabular}
\end{adjustbox}
\caption{Training time of all experimental settings. The total time is obtained by multiplying the per-epoch time with the number of epochs executed, which is 50 in this study. }
\label{table:timing}
\end{table}

\begin{table}[t]
\centering
\begin{adjustbox}{max width=\textwidth}
\begin{tabular}{cccc}
\toprule[.1em]
carotid  & model & time (s/slice) & total time (s/3DUS)\\
\midrule[.1em]
\multirow{4}{*}{CCA} & SDL & 0.050 & 0.786\\
 & SDL$+$TTA & 0.078 & 1.256\\
 & Two-channel U-Net & 0.028 & 0.453\\ 
 & Two-channel U-Net$+$TTA & 0.052 & 0.837\\ 
\midrule[.1em]
\multirow{4}{*}{ICA}  & SDL & 0.054 & 0.542\\
 & SDL$+$TTA & 0.083 & 0.834\\
 & Two-channel U-Net & 0.031 & 0.310\\
 & Two-channel U-Net$+$TTA & 0.056 & 0.563\\
\bottomrule[.1em]\\
\end{tabular}
\end{adjustbox}
\caption{Testing time of all experimental settings. Time required to crop the ICA images is excluded. This time is listed Section \ref{sc:discussion}. -Note that the execution times for two-channel U-Net models using different cost functions (i.e., DDL, TDL and ATDL) are the same. TTA: test time augmentation.}
\label{table:test_timing}
\end{table}

\begin{figure*}[tbp]
	\centering
\renewcommand{\thesubfigure}{a}
\subfloat[]{
\includegraphics[width=0.45\textwidth]{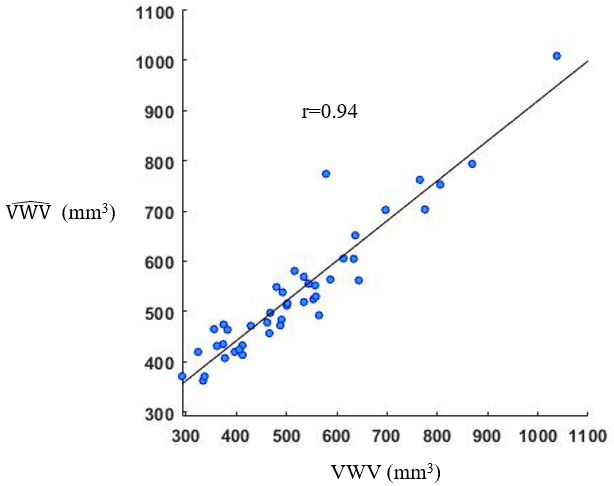}}
\renewcommand{\thesubfigure}{b}
\subfloat[]{
\includegraphics[width=0.5\textwidth]{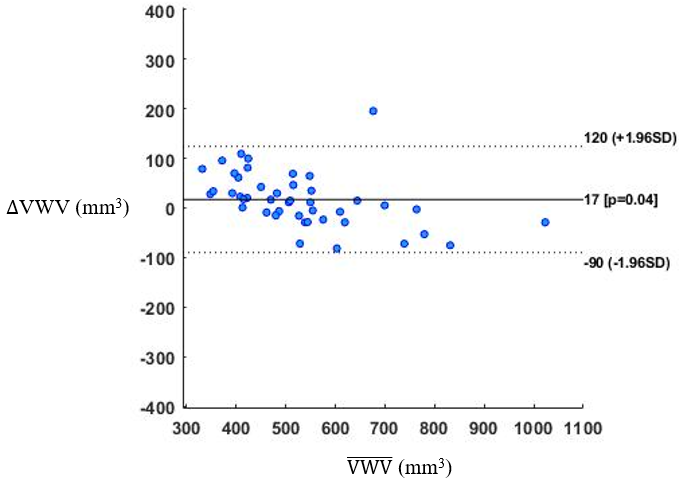}}
	\caption{Relationships between VWV measurements of the CCA obtained from the manually segmented boundaries (VWV) and that obtained algorithm-generated boundaries ($\widehat{VWV}$). (a) The Pearson's correlation and (b) and Bland-Altman analysis of the two sets of measurements.  \label{fig:stats_cca}}
\end{figure*}

\begin{figure*}[tbp]
\centering
\renewcommand{\thesubfigure}{a}
\subfloat[]{
\includegraphics[width=0.45\textwidth]{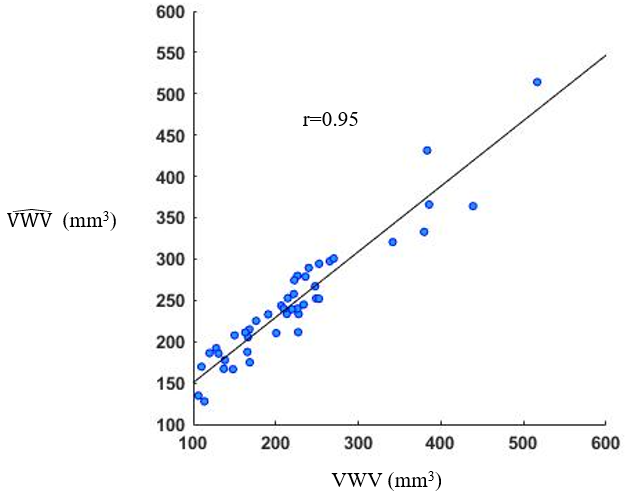}}
\renewcommand{\thesubfigure}{b}
\subfloat[]{
\includegraphics[width=0.5\textwidth]{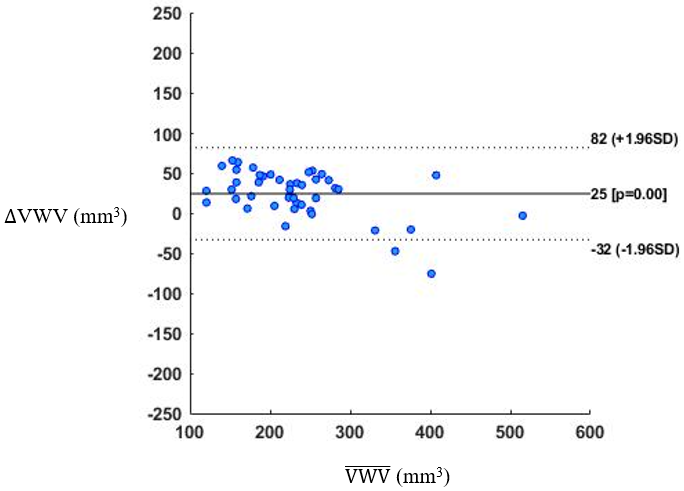}}
	\caption{Relationships between VWV measurements of the ICA obtained from the manually segmented boundaries (VWV) and that obtained algorithm-generated boundaries ($\widehat{VWV}$). (a) The Pearson's correlation and (b) Bland-Altman analysis of the two sets of measurements.  \label{fig:stats_ica}}
\end{figure*}

Fig. \ref{fig:stats_cca} shows the correlation and the Bland-Altman plots comparing the VWV measurements for CCA produced manually and based on the boundaries generated by the ADTL + TTA setting. Fig. \ref{fig:stats_cca} (a) shows a strong and significant correlation between the two VWV measurements ($r=0.94$ and $p=1.22\times10^{-21}$). The Bland-Altman plot shown in Fig. \ref{fig:stats_cca} (b) shows a bias of 17 $mm^3$ with 95\% limits of agreement ranging from -90 to 120 $mm^3$, which is small compared to the VWV measurement range (292 – 1037 $mm^3$). Fig. \ref{fig:stats_ica} shows similar plots for the VWV measurements for the ICA. The Pearson's coefficient is strong and significant ($r=0.95$ and $p=2.53\times10^{-23}$).  The Bland-Altman plot shown in Fig. \ref{fig:stats_ica} (b) shows a bias of 25 mm3 with 95\% limits of agreement ranging from -32 to 82 $mm^3$, which is small compared to the VWV measurement range (106 – 516 $mm^3$).

Tables \ref{table:timing} and \ref{table:test_timing} show the training and testing times required by models evaluated in this study. Since training with SDL to segment MAB and LIB require two single-channel U-Net, the training and testing time of SDL is the summation of computational time required to train and test the independent networks used to segment MAB and LIB. It is not unexpected that testing and training times required by SDL are twice of those required by two-channel U-Nets. Training time required by ATDL is slightly higher than TDL and DDL due to the additional time required to evaluate the ATDL function (Eq. \ref{eq:atdl}). Testing times required for all two-channel U-Net settings (i.e., DDL, TDL and ATDL) are the same, with TTA taking extra evaluation times.

\section{Discussion}
\label{sc:discussion}
Changes of VWV measured from 3D carotid ultrasound images were shown to be sensitive to the effect of medical therapies \cite{krasinski2009three}. Localized vessel wall thickness (VWT) distribution was shown to be even more sensitive than VWV \cite{cheng2017sensitive}. However, the clinical utility of these 3DUS measurements was limited by the requirement for segmenting MAB and LIB manually from 3DUS images, which is time-consuming and subject to observer variability. Although methods with different levels of automation have been previously proposed to segment MAB and LIB from CCA \cite{ukwatta2011three,yang2013ultrasound,ukwatta2013three,menchon2014automatic,hossain2015semiautomatic,azzopardi2017automatic,zhou2019deep,zhou2020voxel,azzopardi2020bimodal}, a substantial gap exists in the automation of ICA segmentation, possibly due to the difficulties in discriminating the internal from the external carotid. However, the VWV and VWT measurements should include ICA as plaque burden is more prevalent in the ICA than CCA \cite{prabhakaran2006carotid}. Therefore, there is a critical requirement for developing an automatic algorithm to segment MAB and LIB from ICA. In this work, we developed an algorithm that (1) automates the segmentation of CCA without ROI identification and (2) segments the ICA based on the ROI identified in the axial slice most proximal and most distal to the bifurcation in the 3DUS volume. Observer interaction was minimal and the interaction time required was 8.7 s per 3DUS volume. We demonstrated that the VWT maps covering the CCA and ICA can be constructed accurately. The proposed method has the potential to accelerate the translation of 3DUS vessel wall measurements to clinical research and clinical practice.
The technical contribution of this paper involves the introduction of the adaptive triple Dice loss (ADTL) function. The TDL function used in a two-channel U-Net to segment the vessel wall in black-blood MRI \cite{wu2019deep}. The authors reported that the two-channel U-Net with TDL function has only a 0.4-0.5\% improvement in Dice coefficients over the conventional U-Net, and conceded that the performance improvement “is not much better” than the conventional U-Net. Our finding is similar. We found that the independent contributions by the two-channel U-Net architecture and the introduction of the TDL function are not statistically significant. ADTL shifts the focus adaptively to the minimization of the Dice loss of vessel wall segmentation in the second half of the training process after the Dice losses are already relatively low at this stage. This adaptive approach has attained a statistically significant improvement over the TDL function. As shown in Fig. \ref{fig:examples}, the improvement is substantial at the bifurcation where the axial cross-section of the artery is elongated.

Table \ref{table:per} compares the DSC attained by our method with previously published methods in CCA segmentation. The DSC of MAB and LIB segmentation produced by our methods is lower than that attained by Ukwatta et al. \cite{ukwatta2011three} and Zhou et al. \cite{zhou2019deep} and the DSC of MAB segmentation was slightly lower than Hossain et al. \cite{hossain2015semiautomatic}. These three methods involve extensive user interactions. Ukwatta et al. \cite{ukwatta2011three} require a user to take 9 minutes to initialize the algorithm. They reduced the initialization time to around 2 minutes in another version of the algorithm \cite{ukwatta2013three}, in which the DSC was lower. The human interaction time was 13.8s in Zhou et al. \cite{zhou2019deep}. The method of Hossain et al. \cite{hossain2015semiautomatic} takes around 14 minutes to initialize. In contrast, no manual ROI identification was needed in our CCA segmentation algorithm. A subtle difference between the evaluation of our algorithm and Zhou et al. \cite{zhou2020voxel} algorithm should be pointed out. The evaluation in Zhou et al. \cite{zhou2020voxel} was done separately for CCA slices far from and close to the bifurcation, which are categorized as the “CCA” and “bifurcation” groups, respectively. No specification as to how close the slice should be for it to be categorized as belonging to the “bifurcation” group. Evaluation was performed separately for these two groups. We list the result for the “CCA” group in Table \ref{table:per}. The DSC in the "bifurcation" group is lower (0.919 vs. 0.932 for MAB and 0.893 vs. 0.895 for LIB). We do not make this differentiation in our evaluation.


\begin{table}[htbp]
\centering
\begin{adjustbox}{max width=\textwidth}
\begin{tabular}{cccc}
\toprule[.2em]
\multirow{2}{*}{Paper} & \multirow{2}{*}{Execution time} & \multicolumn{2}{c}{DSC}\\
& & MAB & LIB\\
\midrule[.2em]
Ukwatta et al. \cite{ukwatta2011three} &  11.1 min & 0.954 & 0.931\\
Yang et al. \cite{yang2013ultrasound} & 4.3 min &  0.944& 0.928 \\
Ukwatta et al. \cite{ukwatta2013three} & 1.72 min & 0.944& 0.906 \\
Hossain et al. \cite{hossain2015semiautomatic} & 37 min & 0.953 & 0.883\\
Zhou et al. \cite{zhou2019deep} & 34.4 s & 0.965 & 0.928\\ 
Zhou et al. \cite{zhou2020voxel} & 0.69 s$^{*}$ & 0.932 & 0.895\\
Proposed algorithm (CCA) & 0.837 s & 0.951 & 0.916\\
\bottomrule[.2em]\\
\end{tabular}
\end{adjustbox}
\caption{Comparison in the DSC and time requirement of CCA segmentation methods from carotid ultrasound. *Reported time excludes the time required to identify ROI }
\label{table:per}
\end{table}

Table \ref{table:per} also compares the time required by different algorithms to segment the CCA in a 3DUS volume. The time tabulated includes the manual initialization time. Many algorithms tabulated involved extensive human interaction \cite{ukwatta2011three,yang2013ultrasound,ukwatta2013three,hossain2015semiautomatic,zhou2019deep}. Therefore, the segmentation time is relatively long. The training time of Zhou et al. \cite{zhou2020voxel} is 11 h, longer than 7 h required by our method (Table IV). The segmentation time required by our approach is shorter than Refs \cite{ukwatta2011three,yang2013ultrasound,ukwatta2013three,hossain2015semiautomatic,zhou2019deep}. Although the CCA segmentation time required by our algorithm is longer than  Zhou et al. \cite{zhou2020voxel} (0.69s/3DUS), their method require manual ROI identification for the CCA, which would take a time comparable to that we took to identify an ROI for ICA (i.e., 8.7s).

Although our method yields promising carotid segmentation performance, manual ROI identification is still needed for ICA segmentation. In the future, a 3D keypoint detection module \cite{simon2017hand} can be implemented as a deep learning network to localize the ICA. An ROI can then be identified automatically to be segmented by the proposed method. This strategy would allow a fully automatic workflow.

\section{Conclusion}
In this paper, we developed a two-channel U-Net trained by the proposed adaptive TDL to segment the MAB and the LIB of the CCA without manual ROI identification and the ICA with minimal user interaction. The proposed approach generates VWV measurement efficiently and accurately for a longitudinal coverage encompassing the CCA and ICA. The  MAB and LIB segmented by the approach can be used to generate VWT maps, from which VWT-based biomarkers can be measured to provide sensitive treatment effect evaluation and accurate stroke risk stratification.

\section{Acknowledgement}
Dr. Chiu is grateful for the funding support from the Research Grant Council of HKSAR, China (Project nos. CityU 11205917, CityU 11203218) and the City University of Hong Kong Strategic Research Grant (Project nos. 7005226 and 7005441). We also thank Dr. J. David Spence for providing the 3D ultrasound images used in this study.
\bibliographystyle{IEEEtran}
\bibliography{report}
\end{document}